\documentclass[aps,prb,twocolumn,epsf,amsmath,showpacs]{revtex4}
\usepackage{epsfig}
\usepackage{bm}
\begin{document}
\title{Magnetic-field symmetries of mesoscopic nonlinear conductance}
\date{\today}
\author{M. L. Polianski}
\email{polian@physics.unige.ch}
\author{M. B\"uttiker}

\address{D\'epartement de Physique Th\'eorique,
Universit\'e de Gen\`eve, CH-1211 Gen\`eve 4, Switzerland.}


\begin{abstract}
We examine contributions to the dc-current of mesoscopic samples
which are non-linear in applied voltage. In the presence of a
magnetic field, the current can be decomposed into components which
are odd (antisymmetric) and even (symmetric) under flux reversal.
For a two-terminal chaotic cavity, these components turn out to be
very sensitive to the strength of the Coulomb interaction and the
asymmetry of the contact conductances. For both two- and
multi-terminal quantum dots we discuss correlations of current
non-linearity in voltage measured at different magnetic fields and
temperatures.

\end{abstract}

\def \esc {\gamma_{\rm esc}}
\def\a{\alpha}
\def\b{\beta}
\def\g{\gamma}
\def\d{\delta}
\def\D{\Delta}
\def \e{\varepsilon}
\def\erg {\tau_{\rm erg}}
\def\h{\hbar}
\def\ph{\varphi}
\def\Tr{\mbox{tr}\,}
\def\P{{\mathcal P}}
\def\S{{\mathcal S}}
\def\U{{\mathcal U}}
\def\l{\lambda}
\def\LL {\Lambda}
\def\m{\mu}
\def\DD{\partial}
\def \lF {\lambda_{\rm F}}
\def\dwell{\tau_{\rm d}}
\def \Diff {{\mathcal D}}
\def\Coop {{\mathcal C}}
\def \WS {Q}
\def\G {{\mathcal G}}
\def \Ga {{\mathcal G}_a}
\def \Gs {{\mathcal G}_s}
\def\A{{\mathcal A}}
\def\la {\langle}
\def\ra {\rangle}
\def \etal{{\em et al.}}
\def \diel {4\pi\epsilon_0\epsilon}
\def \weak {g_{\rm wl}}
\def \Thou{E_{\rm Th}}
\def \as{{\mathcal A}}
\def \bs{\blacksquare}

\maketitle

\section{Introduction}\label{intro}
 Symmetries are of fundamental interest in all fields
of physics. In linear irreversible transport the Onsager-Casimir
\cite{Onsager} symmetry relations are important since they relate
different transport coefficients. Here we are concerned with
electron transport on the mesoscopic scale. This scale emerges when
the distance carriers can travel without losing their phase
coherence becomes comparable to the dimensions of the sample. It has
long been understood that the Onsager-Casimir relations are not
restricted to the macroscopic domain but extend to linear transport
on the mesoscopic scale. For the linear conductance $G_{\alpha\beta}
= dI_{\alpha}/dV_{\beta}$ Onsager-Casimir implies that under field
reversal we have the symmetry $G_{\alpha\beta} (\Phi) =
G_{\beta\alpha}(-\Phi)$. An important condition is that voltages are
measured at contacts which are sufficiently large such that they can
effectively be considered as equilibrium electron
reservoirs\cite{Markus_1986}.

Since symmetries are important it is crucial to understand their
limit of validity. Are Onsager-Casimir relations strictly valid only
in the linear transport regime? What could cause their breakdown? Is
it possible to quantify and measure the departure from symmetry? In
this work we report recent theoretical and experimental progress on
these questions using as an example electron transport through a
chaotic quantum dot. We extend earlier discussions to correlations
of current non-linearity in voltage measured at different magnetic
fields and temperatures. Interestingly, as we will now discuss,
these questions are related to the role Coulomb interactions play in
non-linear transport.

In the mesoscopic regime the linear conductance depends on quantum
interference \cite{meso} and depends on Coulomb interactions. A
change in the Hartree potential is similar to a small change in the
shape of the sample or in its impurity configuration and thus needs
no separate discussion. In contrast Hartree-Fock terms \cite{AA} can
modify the conductance significantly. A well known example of such
an interaction effect is the physics of Coulomb blockade which
becomes relevant at low temperatures if the contacts are
pinched-off. However, if the conductance of the quantum dot with
ballistic contacts is large, $G=(e^2/h)g \gg e^2/h$, Fock terms give
only a small relative correction $\lesssim 1/g$ to sample-specific
quantum fluctuations. In such open dots weak localization
corrections (WL) or universal conductance fluctuations (UCF) remain
universal at low temperatures. In particular, Coulomb interactions
do not affect WL and  just slightly ($\lesssim 1/g$) modify UCF only
at elevated temperatures \cite{BLF}. Since these effects due to
many-body physics in open quantum dots appear only as small
corrections to non-interacting theory, one could conclude that
Coulomb interaction effects are unimportant.

\begin{figure}[t]
\begin{center}
\psfig{file=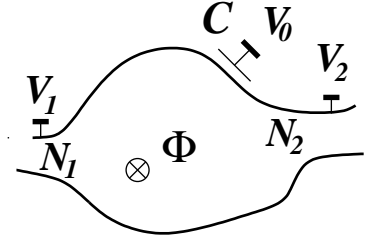,width=5cm} \caption{Quantum dot with magnetic
flux $\Phi$ and dc bias voltages $V_{1,2}$ at the contacts and $V_0$
at the gates with capacitance $C$.}\label{fig1a}
\end{center}
\end{figure}

However, Coulomb interactions play a much more important role in
non-linear transport. Experimentally transport is characterized by
the {\em full} conductance $I/V$ rather then the differential
conductance, $dI/dV\neq I/V$. In particular, here we are interested
in $(1/2)\DD^2 I/\DD V^2$ (at $V = 0$) which we call the
second-order non-linear conductance or simply the non-linear
conductance. In this regime it is the Coulomb interaction that
breaks the symmetry of the Onsager-Casismir relations. Conversely,
measurement of transport non-linearity provides a tool to determine
interactions.

Initial discussions of non-linear dc- transport \cite{AK,KL} and the
rectification of an ac-applied voltage \cite{FK} in mesoscopic
samples were addressed without taking interaction effects into
account. Interestingly, inclusion of Coulomb interaction does not
only modify these phenomena but also leads to a qualitatively new
effect: namely, even in a two terminal conductor it is possible to
have a component to the current which is odd in magnetic field
$I(\Phi) \neq I(-\Phi)$. In generic mesoscopic conductors, in
chaotic cavities (quantum dots) and in metallic diffusive conductors
this effect vanishes when an average is taken over many samples of
slightly different geometry or impurity configuration (mesoscopic
averaging) \cite{SB,SZ}. Thus in generic mesoscopic conductors this
is purely a quantum effect.

S\'anches and B\"uttiker investigated the effect of interactions on
current-voltage characteristic of open chaotic (ballistic or
diffusive) cavities, see Fig. 1, in the presence of a magnetic flux
$\Phi$ comparable to a flux quantum $\Phi_0=eh/c$. \cite{SB} They
found the dependence of the fluctuation of the antisymmetric
component on the interaction strength and the dependence on the
number of ballistic channels transmitted through the contacts of the
cavity. Spivak and Zyusin explored this asymmetry at small fields,
$\Phi\ll \Phi_0$, and weak interactions in open diffusive samples
\cite{SZ}. Under these conditions the asymmetric fluctuations are
linear in flux and interaction strength. Polianski and B\"uttiker
\cite{PB} present a theory which describes the entire crossover from
the regime of low magnetic fields to the regime considered in Ref.
\onlinecite{SB} in which the flux through the sample is comparable
to a flux quantum $\Phi_0$.

\begin{figure}[t]
\begin{center}
\psfig{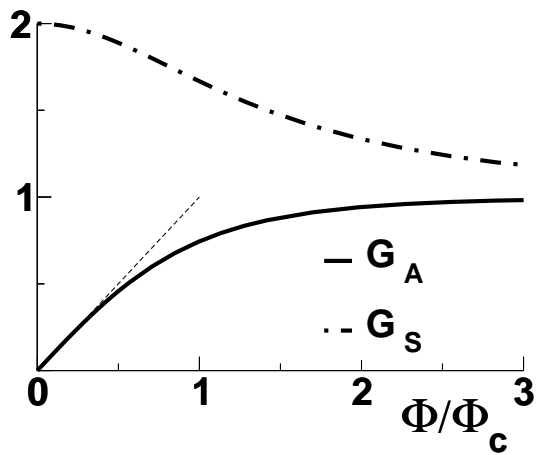} \caption{Normalized fluctuation
of the (anti)symmetric nonlinear conductance $({\mathcal
G}_{a}){\mathcal G}_{s}$ as a function of flux
$\Phi/\Phi_c,\Phi_c\ll\Phi_0$ for strongly interacting dot with
symmetric contacts, from \cite{PB}.}\label{fig1b}
\end{center}
\end{figure}

The experiments measured field asymmetry in a wide range of
structures: in nanotubes \cite{wei}, ballistic billiards
\cite{marlow}, ballistic quantum dots \cite{Zumbuhl}, ballistic
\cite{ensslin} and diffusive \cite{Bouchiat} Aharonov-Bohm rings.
This motivated further research on non-linear quantum
\cite{BS,Tsvelik} and classical \cite{AG} effects. The experiment by
Zumb\"uhl \etal \cite{Zumbuhl} investigated this asymmetry in
chaotic ballistic dots at very low voltages. The experiment
demonstrated that the asymmetry vanishes on average and depends
linearly on flux at sufficiently small fields. The experiment also
investigates the dependence of the asymmetry on the number of
channels in the contacts of the sample. Some experimental results,
especially, the magnitude of the asymmetry, if four or more channels
are open, agree well with existing theories \cite{SB,SZ,PB}. A more
detailed investigation of the dependence on various external
parameters is still necessary.

Theoretically Ref. \onlinecite{PB} considers fluctuations of {\em
both} anti-symmetric and symmetric component of non-linear
conductance
\begin{eqnarray}\label{eq:1defGaGs}
\binom{{\mathcal G}_{s}(\Phi)}{{\mathcal
G}_{a}(\Phi)}\equiv\frac{h}{4e^2}\left.\frac{\DD^2}{\DD
V^2}\left(\frac{I(\Phi)\pm I(-\Phi)}{2}\right)\right|_{V\to 0}
\end{eqnarray}
through a 2-terminal quantum dot with arbitrary interaction
strength, flux, temperature and dephasing \cite{PB}. The
fluctuations (square root of the variance) of these two quantities
are shown in Fig. \ref{fig1b} as a function of the flux applied to
the cavity. Interestingly, similarly to linear transport (see
references in \cite{Beenakker}), the crossover of $\Ga$ and $\Gs$
from low to high magnetic fields is shown to depend on the flux
$\Phi_c\ll\Phi_0$ and not on the flux quantum $\Phi_0$
characteristic for open diffusive samples. Since the dot is
essentially zero-dimensional, the dwell time $\dwell$ an electron
spends inside is much larger then the ergodic time $\erg$ necessary
to explore the phase space of the dot. Therefore an electronic
trajectory gains a flux $\sim \Phi_0$ during $\dwell/\erg\gg 1$
random explorations of the dot already at $\Phi\sim
\Phi_0/\sqrt{\dwell/\erg}\ll\Phi_0$. Fluctuations of $\Ga,\Gs$ occur
on the background of an ensemble averaged anti-symettrized $\la \Ga
\ra=0$ and symmetrized $\la \Gs\ra=0$ non-linear conductance. The
magnitude of these fluctuations as a function of flux $\Phi/\Phi_c$
is shown in Fig. \ref{fig1b} for the limit of strongly interacting
dot with symmetric contacts (corresponding to the experiment
\cite{Zumbuhl}).

However, if either the Coulomb interaction is not very strong or
ballistic contacts are unequal, $N_1\neq N_2$ on Fig. \ref{fig1a},
the field-asymmetry of the non-linear signal may be strongly reduced
compared to strongly interacting symmetric set-up. In practice, to
observe a strong violation of Onsager relations
\cite{wei,marlow,Zumbuhl,ensslin,Bouchiat} one might want to
consider not the amplitude of $\Ga$, but rather the relative
asymmetry in the signal, denoted here ${\A}=\Ga/\Gs$. Below we
discuss the optimal regime to maximize this ratio. We find how
magnetic field and temperature affect fluctuations of this asymmetry
parameter and find the contact asymmetry that maximizes $\A$. For
sufficiently strong interaction, symmetric contacts turn out to be
the optimal regime. Surprisingly, however, if the interactions are
not very strong an asymmetric set-up, $N_1\neq N_2$, is more
advantageous.

Since both field-components $\Ga, \Gs$ were recently explored
experimentally \cite{Zumbuhl,Bouchiat}, we consider the statistics
of $\Ga$ and $\Gs$ and their correlations for arbitrary
interactions. Particularly, in the non-interacting limit we find an
universal relation between UCF and fluctuations of $\Gs$,
independent of temperature and applied field. In view of a
multi-terminal experiment \cite{ensslin}, we generalize our
treatment of an asymmetric component $\Ga$ to a multi-terminal dot
and consider non-linearity not only with respect to a contact, but
also to the gate voltage. In the end we discuss interaction effects
in linear vs. non-linear transport, applicability of existing
theories to experimental data and provide a formula for fluctuations
of the asymmetry $\A$ in the experimental regime of Ref.
\onlinecite{Zumbuhl}.

The paper is organized as follows: Sec. \ref{model} describes the
model and the basic steps of our approach. We present our results in
Sec. \ref{results} and discuss them in Sec. \ref{discussion}.

\section{Model}\label{model}
The 2D chaotic quantum dot, see Fig. \ref{fig1a}, is biased with dc
voltages $V_{\a},\a=1,...,M$ at contacts with $N_\a$ ballistic
channels, and by the voltage $V_0$ at the gates with capacitance
$C$. Chaos in the dot is either due to diffusive scattering off
impurities, when the mean free path is smaller then the size of the
dot, $l\ll L$, or due to the irregular shape of a ballistic dot with
chaotic classical dynamics, $l\gg L$. The dot is in the universal
regime \cite{Beenakker,ABG}, when the Thouless energy $E_{\rm
T}=\hbar/\erg$ is the largest energy-scale in the problem. Typically
$\Thou\sim \hbar \mbox{ min }\{ lv_{\rm F}/L^2, v_{\rm F}/L\}$, see
the review \cite{Beenakker} for details. The mean level spacing (per
spin direction) $\Delta=2\pi\hbar^2/(m^*\mbox{ Area})$ and the
number of conducting channels $N$ together define the dwell time
$\dwell=h/(N\Delta)\gg \erg$ which an electron typically spends
inside the dot.

Scattering is assumed spin-independent and this spin degeneracy is
explicitly accounted for by the factor $\nu_s$. The number $N$ of
ballistic (orbital) channels in the contacts is assumed to be large,
$N\gg 1$, so that the problem can be considered analytically. We use
Random Matrix Theory (RMT) for the energy-dependent scattering
matrix $\S(\e)$ and refer reader to reviews \cite{Beenakker,ABG} for
details of RMT and its relation to Green function technique. The
diagrammatic technique based on small parameter $1/N\ll 1$ expansion
are given in \cite{diagram} for energy-independent scattering matrix
$\S$ and in \cite{iop} for energy-dependent $\S(\e)$. We also
require that the transport is only weakly nonlinear and refer to
\cite{KL,LBM} for a discussion of highly nonlinear transport in
mesoscopic samples. Here we assume that $eV\ll N\Delta$ and treat
the $I-V$ nonlinearity only to order $(eV)^2$, see Eq.
(\ref{eq:1defGaGs}). The energy dependence of the $\S$-matrix allows
us to consider non-zero temperatures $T\ll \Thou$, and for
convenience we normalize it to dimensionless parameter $t\equiv 2\pi
T/(N\Delta)$.

The magnetic field appears via the total flux $\Phi$ through the
dot. For simplicity the dot is assumed here to be (nearly) circular
with radius $L$. As discussed in Sec. \ref{intro}, the relevant
flux-scale for the dot in the diffusive and the ballistic regime is
$\Phi_c\ll\Phi_0$ \cite{Beenakker,PB}. Therefore for convenience we
normalize magnetic flux $\Phi$ through the dot to dimensionless flux
$\phi$ as follows:
\begin{eqnarray}\label{eq:Flux}
\phi=\frac{\Phi}{\Phi_0} \frac{\sqrt{\dwell v_F l}}{2L}.
\end{eqnarray}
To describe the crossover from the diffusive to the ballistic regime
we use the substitute $l\to \pi L/4$  according to \cite{Adam} (the
numerical factor given in \cite{FrahmPichard,Beenakker} is not
correct). The importance of crossover is evident if one notices that
much of the data \cite{Zumbuhl} taken pertains to this intermediate
regime.

The geometrical capacitance $C$ of the dot can be defined, {\em
e.g.} in the constant interaction model in the Hamiltonian approach
\cite{ABG} from the solution of electrostatic problem, if the shape
of the dot is known exactly and all nearby conductors are taken into
account. Usually such an exact solution is not available, but this
capacitance can be estimated either as $C\sim \epsilon L$ for an
ungated sample or as $C\sim \epsilon L^2/d$ for macroscopic gates
separated by a distance $d$ from the dot. (Here $\epsilon$ is the
dielectric constant). This capacitance defines the strength of the
Coulomb interaction, and electronic repulsion is referred here as
'strong' if a typical charging energy is large compared to the level
spacing, $e^2/C\gg\Delta$.

\begin{figure}[t]
\begin{center}
\psfig{file=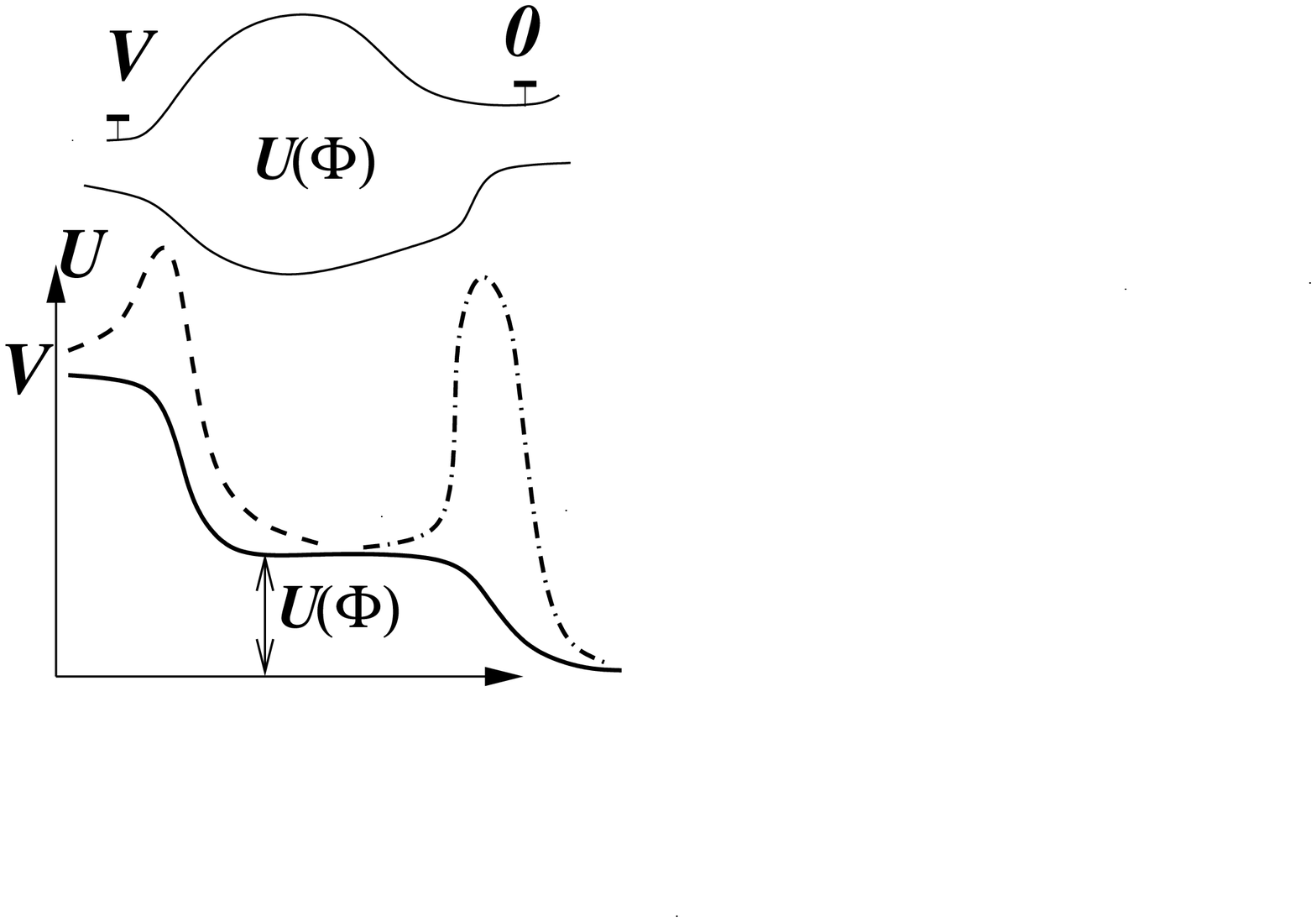,width=9cm}
\caption{Electric potential $U(x)$ shifts the bottom of the energy
band. Bold curve corresponds to open channels and dashed
(dot-dashed) correspond to (classically) reflected channels.
Potential $U$ varies only in the contact regions and remains
constant inside the dot.} \label{fig2}
\end{center}
\end{figure}

For strong screening in the dot, $r_s=(k_{\rm
F}a_B)^{-1}=e^2/(\epsilon\hbar v_{\rm F})\lesssim 1$, an RPA
treatment of Coulomb interactions is sufficient. Furthermore, for
large dots, $L\gg a_B$, only the long distance screening is
relevant. As a consequence, electrons with kinetic energy $\e$ have
a well-defined electro-chemical potential $\tilde \e_\a=\e-eV_\a$ in
the contact $\a$ and $\tilde\e=\e-eU(\vec r)$ in the dot.
For a quantum dot large compared to the Bohr radius but still so
small that its dimensionless conductance, $g_{\rm dot}=\Thou/\Delta$
is much larger then conductance of the contacts, $g_{\rm dot}\gg N$,
the potential can be taken uniform ('zero-mode
approximation')\cite{ABG}. The leading interactions are then present
in the form of a Hartree electrical potential $U$ which shifts the
bottom of the energy band in the dot, see Fig. \ref{fig2} and
therefore modifies the $\S$-matrix.

Therefore, transport depends on the Fermi-distributions
$f(\tilde\e_\a)$ and the scattering matrix $\S(\tilde \e)$. The
current in contact $\a$ is $I_\a=\int d\e I_\a(\e)$ and for $eV\ll
N\Delta$ the spectral current $I_\a(\e)$ can be expanded in powers
of $eV$:
\begin{eqnarray}\label{eq:I_I}
&&I_\a (\e)=\frac{\nu_s e^2}{h}
\sum_{\d=1}^M f(\tilde\e_\d)
\Tr \left[{1\!\! 1}_\a\d_{\a\d}-{1\!\! 1}_\d
\S^\dagger(\tilde\e){1\!\! 1}_\a
\S(\tilde\e)\right]\nonumber\\
&&\approx \frac{\nu_s e^2}{h}\sum_{\b} g_{\a\b}(\e)V_\b+\frac{\nu_s
e^3}{h}\sum_{\b\g}g_{\a\b\g}(\e)V_\b V_\g.
\end{eqnarray}
In Eq. (\ref{eq:I_I}) the total current $I_\a$ is expressed in terms
of the dimensionless linear conductance at energy $\e$,
$g_{\a\b}(\e)=\Tr({1\!\! 1}_\a\d_{\a\b}-{1\!\! 1}_\b
\S^\dagger(\e){1\!\! 1}_\a \S(\e))$ and the nonlinear conductance
(measured in units of inverse energy) $g_{\a\b\g}(\e)=(h/2\nu_s
e^3)\DD^2 I_\a(\e)/\DD V_\b\DD V_\g$, which depends on the potential
$U$. To this accuracy $U$ needs to be known only up to the first
order derivatives, the characteristic potentials $u_\a=\DD U/\DD
V_\a$ \cite{mb93,ChristenButtiker,Christen97}. The characteristic
potentials $u_\a$ are found self-consistently \cite{RPA} from
current conservation and gauge-invariance requirements. The dc limit
$\omega\to 0$ of an ac-result \cite{iopBP} reads
\begin{eqnarray}\label{eq:u}
u_\a &=&\frac{-\int d\e f'(\e)\Tr \S^\dagger \DD_\e\S {1\!\!
1}_\a}{2\pi i C/\nu_s e^2 -\int d\e f'(\e)\Tr \S^\dagger
\DD_\e\S},\\
u_0 &=& \frac{2\pi i C/\nu_s e^2}{2\pi i C/\nu_s e^2 -\int d\e
f'(\e)\Tr \S^\dagger \DD_\e\S}=1-\sum_{\a=1}^M u_\a.\label{eq:u0}
\end{eqnarray}
We point out that this self-consistent treatment of Coulomb
interactions has certain limitations. Remaining in a single-particle
picture of scattering, we neglect many-body effects. We consider
here a dot with multi-channel ballistic contacts, $N\gg 1$ when
scattered wave-packets are delocalized. Then the effects of Fock
terms are small $\lesssim 1/N$ compared to that of Hartree
self-consistent potential \cite{BLF,VA}.

The sample-specific fluctuations of $u_\a$ are dependent on magnetic
flux $\phi$, and determine the field-asymmetry of the non-linear
conductance. These derivatives are used to express the conductances
 $g_{\a\b\g}(\e)$:
\begin{eqnarray}\label{eq:Gabg}
g_{\a\b\g}(\e)=\frac{-f'(\e)}{2} \left(\d_{\b\g} g'_{\a\b}-u_\b
g'_{\a\g}-u_\g g'_{\a\b}\right),
\end{eqnarray}
where the prime stands for energy derivative. Transport coefficients
in general depend on magnetic field due to sensitivity of the matrix
$\mathcal S$ to magnetic field. For example, field inversion
transposes the scattering matrix $\S(\phi)=\S^{\rm T}(-\phi)$. The
response to the gate voltage remains invariant under such inversion,
$u_0(-\phi)=u_0(\phi)$, but the response to the voltage at the
contacts is asymmetric, $u_\a(\phi)\neq u_\a(-\phi)$.

We point out that this $\phi$-sensitivity of electrostatic potential
$U(\vec r,\phi)$, which locally shifts the bottom of the energy
band, is a general feature of coherent systems. In the high-field
limit of Quantum Hall bar a voltage applied to one of the contacts
changes the electric potential only in the outgoing edge, but not in
the incoming. The field inversion reverses the direction of edge
channels, so the same voltage induces a potential on the opposite
edge, $U(\vec r,\Phi)\neq U(\vec r,-\Phi)$. The magnetic field
asymmetry in an edge state geometry with a Coulomb blockaded
impurity is the subject of Ref. \onlinecite{SB_Coulomb}. Even at the
small fields, $\Phi\lesssim \Phi_0$, of interest here, the potential
$U(\vec r)$ remains field-sensitive, although the field effect is
not as drastic as in the edge state geometries mentioned above.

Since the dot is a good conductor compared to contacts,
$\Thou/\Delta\gg N$, the voltage drops mainly over the ballistic
contacts and remains uniform inside the dot. This is qualitatively
sketched by the bold curve in Fig. \ref{fig2} (dashed curves
correspond to the potential for channels reflected by the ballistic
constriction). Therefore in a quantum dot we deal with a uniform
potential $U(\vec r,\phi)=U(\phi)$. This is very different from an
open diffusive sample, where $\erg\sim\dwell$ and the voltage drops
gradually.

Classically the voltage through such a conducting nod is divided
according to the widths of the contacts, $u_{{\rm cl }\,\a}=\DD
U_{\rm cl}/\DD V_\a=N_\a/N\sim 1$, is insensitive to magnetic field.
However, since the phases of scattering amplitudes between different
channels are modified by the magnetic field, a field-sensitive and
sample-specific correction $\d u_\a=u_\a-u_{{\rm cl }\,\a}\sim 1/N$
appears. Although at such small fields it is small compared to
classical the value, its asymmetry with respect to magnetic field
$\d u_\a(\phi)\neq \d u_\a(-\phi)$ leads to the asymmetry in the
current-voltage characteristic, and more specifically in the
non-linear conductance $g_{\a\b\g}(\phi)\neq g_{\a\b\g}(-\phi)$.
Below, these (anti-)symmetric components of conductance, $({\mathcal
G}_{a}){\mathcal G}_{s}$,
\begin{eqnarray}\label{eq:defGaGs}
\binom{{\mathcal G}_{s}}{{\mathcal G}_{a}}_{\a\b\g}\equiv\frac
12\int d\e\left(g_{\a\b\g}(\e,\phi)\pm g_{\a\b\g}(\e,-\phi)\right),
\end{eqnarray}
are investigated in detail (for a 2-terminal dot Eq.
(\ref{eq:defGaGs}) corresponds to Eq. (\ref{eq:1defGaGs}) for
$\a=\b=\g=1$).
\section{Results}\label{results}
We group our results into two subsections. First in Sec. \ref{2t} we
consider two-terminal dots and discuss properties of the ensemble
averaged non-linear conductances $\Ga,\Gs$. Since the
field-asymmetry of the measured signal is characterized by the
relative strength of the components introduced in Eq.
(\ref{eq:defGaGs}), we start by considering fluctuations and
correlations of $\A=\Ga/\Gs$, and discuss the dependence of
$\A,\Ga,\Gs$ on contact width and arbitrary interactions. Second, in
Sec. \ref{multit}, we generalize our treatment of $\Ga$ to
multi-terminal quantum dots and discuss the role of a gate voltage.

\subsection{Two-terminal dots}\label{2t}
Here we consider the (anti-) symmetric nonlinear conductance in
two-terminal quantum dots in terms of the $\S$-matrix and its energy
derivative $\partial_\e \S$. To be definite, we set $V_2=0$ and
consider derivatives with respect to $V_1$ only (the gate voltage
$V_0$ is considered in Sec. \ref{multit}). With this set of
voltages, the non-linear conductances are ${\mathcal G}_{a(s)}
\equiv {\mathcal G}_{{a(s)},111}$. One convenient representation of
$\G_{a,s}$ is given in terms of a traceless matrix $\Lambda$, which
is often used for the (dimensionless) linear conductance $g$ of a
quantum dot
\begin{eqnarray}\label{eq:linear}
g= \frac{N_1N_2}{N}-\Tr \Lambda\S^\dagger\Lambda\S
,\,\,\,\Lambda\equiv \frac{N_2{1\!\! 1}_1-N_1{1\!\! 1}_2}{N}
\end{eqnarray}
to separate the classical non-fluctuating part from the quantum
contribution, the second term in Eq. (\ref{eq:linear}). The
projection operator ${1\!\! 1}_\a$ in Eq.(\ref{eq:linear})
corresponds to the diagonal matrix with a unit block for channels of
the $\a$-th lead and zero otherwise. Another equivalent way is to
use the (real) injectivity $\DD_\e \bar n_\a$ and the emissivity
$\DD_\e \underline n_\a$ of the dot into and out of the contact $\a$
\cite{mb93,Christen97}:
\begin{eqnarray}\label{eq:inj_emis}
\DD_\e\bar n_\a &\equiv&\frac{1}{2\pi i}\Tr  {1\!\! 1}_\a\S^\dagger
\DD_\e\S,\\
\DD_\e{\underline n}_\a &\equiv &\frac{1}{2\pi i}\Tr
\S^\dagger{1\!\! 1}_\a\DD_\e\S.
\end{eqnarray}
Using the matrix $\Lambda$ or Eq. (\ref{eq:inj_emis}) we factorize
$\Ga,\Gs$ into products of fluctuating quantities:
\begin{eqnarray}\label{eq:defG}
{\mathcal G}_{a,s}&=&\frac{\pi}{\Delta^2}\frac{\int\int d\e
d\e'f'(\e)f'(\e')\chi_1(\e)\chi_{2,a(s)}(\e')}{C/(e^2\nu_s)-\int d\e
f'(\e)\Tr \S^\dagger \DD_\e\S/(2\pi i)},\\
\chi_1(\e) &=& \frac{\Delta\DD_\e }{2\pi}\Tr\Lambda {\mathcal
S}^\dagger\Lambda{\mathcal S},\label{eq:chi1}\\
\chi_{2,a}(\e) &=&\frac{\Delta}{2\pi i}
\Tr\Lambda[\DD_\e\S,\S^\dagger]=\Delta\left( \DD_\e{\underline
n}_1-\DD_\e\bar n_1\right)
\label{eq:chi2a},\\
 \chi_{2,s}(\e) &=&\frac{\Delta}{2\pi i}\Tr ({1\!\! 1}_2\S^\dagger-\S^\dagger {1\!\! 1}_1)\DD_\e\S
 +\frac{C\Delta}{e^2\nu_s}\nonumber \\ &=&\Delta\left(\DD_\e\bar
 n_2-\DD_\e {\underline n}_1+\frac{C}{e^2\nu_s}\right).\label{eq:chi2s}
\end{eqnarray}
The injectivity and emissivity are related via $\Sigma_\a \DD_\e
\bar n_\a =\Sigma_\a \DD_\e \underline n_\a=\Tr \S^\dagger
\DD_\e\S/(2\pi i)$ and $\DD_\e\bar n_\a(\phi)= \DD_\e{\underline
n}_\a(-\phi)$. Their mesoscopic averages $\la \DD_\e \bar
n_\a\ra=\la\DD_\e \underline n_\a\ra=N_\a/(N\Delta)$ are defined by
the coupling of the dot to the contact $\a$. In contrast to their
classical values, there are quantum fluctuations that render $\DD_\e
\bar n_\a \neq \DD_\e\underline n_\a$ and thus lead to non-zero
r.h.s. in Eq. (\ref{eq:chi2a}). These averages readily demonstrate
that $\la\chi_{2a}\ra=0$, but $\la\chi_{2s}\ra=C\Delta/(\nu_s
e^2)+(N_2-N_1)/N$.

One can also demonstrate that the mesoscopically averaged
$\la{\mathcal G}_{a}\ra,\la{\mathcal G}_{s}\ra=0$ using averaging
over an ensemble of energy-dependent matrices (which are also
symmetric in the absence of a magnetic field). Either component of
$\G$ is expressed by Eq. (\ref{eq:defG}) as a combination of
fluctuating quantities, the most important being fluctuations of the
numerator. One can notice from Eqs. (\ref{eq:linear}, \ref{eq:chi1})
that $\chi_1(\e)\propto \DD_\e G(\e)$. The latter was considered and
proven to vanish upon ensemble averaging in Ref. \onlinecite{deriv}.
This leads to a vanishing of the ensemble average of the non-linear
conductance $\la{\mathcal G}_{a}\ra,\la{\mathcal G}_{s}\ra=0$.
Consequently the magnetic field asymmetry discussed here is a
signature of a quantum effect, similarly to {\em e.g.} the dc
current generated by an ac voltage \cite{FK}. (Interestingly, also
quantum pumping \cite{Brouwerpump} exhibits a magnetic field
asymmetry \cite{Moskalets} similar to the one discussed here.) Thus
the magnitudes of $\Ga,\Gs$ have to be described in terms of
fluctuations or pair-correlation functions averaged over a relevant
ensemble of $\S$-matrices.

Averaging of $\S(\e)$ is performed using the diagrammatic technique of an expansion
in $1/N\ll 1$ \cite{diagram,iop}. For ballistic contacts the results
greatly simplify, since energy- and flux-dependent elements of
the scattering matrix vanish, $\la\S_{ij}(\e,\phi)\ra=0$, and the
non-vanishing pair correlator reads
\begin{eqnarray}\label{eq:pair}
\la\S_{ij}(\e,\phi)\S_{kl}^*(\e',\phi')\ra
=\d_{ik}\d_{jl}\Diff_{\e-\e'}+
\d_{il}\d_{jk}\Coop_{\e-\e'},&&\\
\label{eq:channels} \binom{{\mathcal C}_{\e-\e'}}{{\mathcal
D}_{\e-\e'}} =
\frac{1}{N}\cdot\frac{1}{1+(\phi\pm\phi')^2-2\pi i(\e-\e')/N\Delta}. &&
\end{eqnarray}
The 4th-order correlator can be also expressed in terms of the
Cooperon $\Coop_{\e-\e'}(\phi,\phi')$ and the Diffuson
$\Diff_{\e-\e'}(\phi,\phi')$ (higher-order correlators are presented
in Refs. \onlinecite{BLF,Brouwer_Rahav}). We point out that in the
Green function technique with random Hamiltonian the correlators
$\la G_{ij}^{\rm R}(\e,\phi)G_{lk}^{\rm A}(\e',\phi')\ra$ are given
by Eq. (\ref{eq:pair}) (up to a normalization factor) \cite{ABG}.

The denominator of Eq. (\ref{eq:defG}) is a self-averaging quantity,
$\langle(...)^2\rangle=\langle (...)\rangle^2=(C/(C_\m\Delta))^2$,
with the electrochemical capacitance $C_\m\equiv C/(1+C\Delta/(\nu_s
e^2))$ \cite{PietMarkus} ($\Delta/\nu_s$ stands for the mean level
spacing of spin-degenerate system). The crossover from weak to strong
interaction corresponds to the increase of $C_\m/C$ from
$C_\mu/C\to\nu_s e^2/C\Delta\approx 0$ to 1. The diagrammatic technique
for matrices $\S(\e)$ proves that the functions $\chi_1(\e,\phi)$
and $\chi_{2a, 2s}(\e',\phi')$ are mutually uncorrelated, so that
only correlators $\la\chi_{2s}^2,\chi_{2a}^2\ra$ are important here.

The simplest quantity of interest is the relative asymmetry, the
ratio $\A\equiv \Ga/\Gs$,
 which strongly depends on the
interplay between the interaction strength and the asymmetry of the
contacts. In an experiment it may be desirable to maximize the
magnitude of $\A$. This quantity depends only on $\chi_{2a,2s}$ and
vanishes on the average, $\la {\A}\ra=0$. Its correlations are
non-trivial and as functions of temperature $t,t'=(2\pi/N\Delta)
T,T'$ and flux $\phi,\phi'$ defined in Eq. (\ref{eq:Flux}) they are
\begin{eqnarray}\label{eq:AA}
\la{\A}(t,\phi) {\A}(t',\phi')\ra &=&\frac{{\mathcal F}_-}{{\mathcal
F}_++\frac{N^2}{2N_1N_2}
\left(\frac{C}{C_\mu}-\frac{2N_1}{N}\right)^2},
\\
{\mathcal F}_\pm ={\mathcal F}_\Diff\pm {\mathcal F}_\Coop
&=&\frac{1}{N^2} \int_0^\infty \frac{\pi tx}{\sinh \pi tx}\frac{\pi
t'x}{\sinh \pi t'x}e^{-x}dx \nonumber
\\ &\times &
\left(\frac{e^{-(\phi-\phi')^2 x}}{1+(\phi-\phi')^2}\pm
\frac{e^{-(\phi+\phi')^2 x}}{1+(\phi+\phi')^2}\right) .
\label{eq:Fraw} \nonumber
\end{eqnarray}
Note that due to interactions $C_\mu/C \neq 0$ the expression
(\ref{eq:AA}) is not symmetric with respect to $N_1\leftrightarrow
N_2$ which reflects the fact that we deal with a non-equilibrium
transport coefficient.

First we consider {\em high-field limit} $|\phi+\phi'|\gg 1\gg
|\phi-\phi'|$,  when the Cooperon contribution vanishes ${\mathcal
F}_\Coop\ll {\mathcal F}_\Diff$ so that ${\mathcal F}_-\to{\mathcal
F}_+$. In case of strong interactions in a dot with symmetric
contacts $C/C_\mu-2N_1/N\to 0$, so that
$\la\A(t,\phi)\A(t',\phi')\ra\to 1$ and $\Ga,\Gs$ could be plotted
on the same scale. see Fig. \ref{fig1b}. If either interactions
become weak {\em or} the contacts become strongly asymmetric, the
magnetic-field asymmetry vanishes, ${\A}\to 0$.

For strong interactions even a small asymmetry
in the multi-channel contacts
$N_1=N/2\pm 1$ can make the second term in the denominator of Eq.
(\ref{eq:AA}) dominant and ${\A}\to 0$. The important role of contact
asymmetry becomes clear
 from Eqs. (\ref{eq:chi2a},\ref{eq:chi2s}): quantum
 fluctuations $\d\chi_2\sim \mbox{min }
 \{1, \sqrt{1/t}\}/N$ of numerator and denominator of
 $\A\sim\chi_{2a}/\chi_{2s}$
 occur on the background of $\la\chi_{2a}\ra=0$
 and
 $\la\chi_{2s}\ra=C/C_\mu-2N_1/N$. The numerator grows
 proportionally to $\d\chi_{2}$. In contrast for
 the denominator the fluctuations can be totally neglected if
 its classical average value is large $\la\chi_{2s}\ra\gg \d\chi_{2}$
 and therefore ${\A}\ll 1$.
\begin{figure}[t]
\begin{center}
\psfig{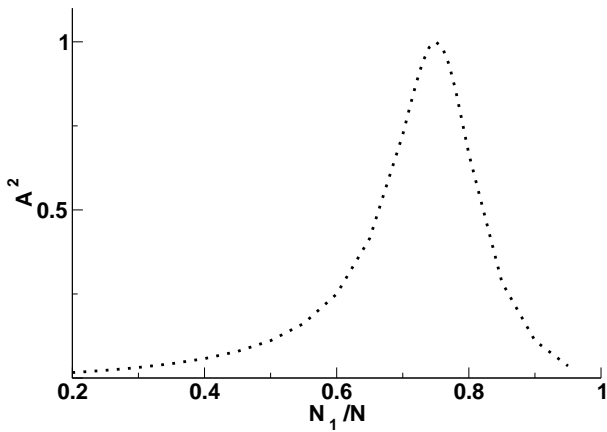}
\caption{Variance of ${\A}=\Ga/\Gs$ as a function of contact
asymmetry reaches maximum at $N_1/N=C /2C_\mu$ (presented for $N=4$
and $C/C_\mu=1.5$). As $N\to\infty$ maximum becomes sharper}
\label{fig3}
\end{center}
\end{figure}

In an experiment the number of channels can be adjusted to maximize
 asymmetry $\A$. In a weakly
interacting dot with $C/C_\mu>2$ the maximum of ${\A}^2 $ is reached
at $N_1/N=C/2(C-C_\mu)$, but it is still small ${\A}\ll 1$ due to
the weak interactions. If the interaction is relatively strong,
$C/C_\mu< 2$ then at $N_1/N=C/2C_\mu$ the maximum ${\A}=1$ is
achieved, see Fig. \ref{fig3}. It is this contact width that
minimizes the magnitude of $\la\chi_{2s}\ra$.

In a realistic quantum dot with symmetric contacts \cite{Zumbuhl}
the interaction is reasonably strong and {\em at intermediate
magnetic fields $|\phi\pm\phi'|< 1$} we have ${\A}^2={\mathcal
F}_-/{\mathcal F}_+$. However, it is important to emphasize that if
a dot becomes smaller, the ratio $e^2/C\Delta$ ($\propto L/a_B$ for
a dot without gates) can diminish as well. Therefore one could
expect that interactions are not necessarily strong and at
$1<C/C_\mu<2$ it would be important to make the contacts asymmetric
to clearly observe a magnetic-field asymmetry. The peak in the Fig.
\ref{fig3} becomes narrower as $N\to\infty$ and the choice of
contact width such that $N_1/N\approx C/2C_\mu$ becomes crucial.
This behavior is a consequence of the scaling of the numerator of
Eq. (\ref{eq:AA}): the width of this peak as a function of $N_1/N$
is proportional to fluctuations $\d\chi_2$ discussed above. Although
experiments are often performed with symmetric contacts to avoid
parasitic effects of an asymmetric circuit, our results suggest that
using contact asymmetry can {\em enhance} the magnetic-field
asymmetry.

Fluctuations of $\A$ can not be found in closed form, since
${\mathcal F}_\pm$ are complicated functions of $t$ and $\phi$, see
Eq. (\ref{eq:AA}). At low temperatures $t\ll 1$ the
temperature-dependence of ${\mathcal F}_\pm$ can be neglected. In
the regime $t\gtrsim 1$, more relevant for experiment \cite{Zumbuhl}
(see also discussion in Sec. \ref{discussion}), the fluctuations of
$\A$ for a dot with symmetric contacts $N_1=N_2$ can be expressed in
 terms of linear conductance $g(\phi)$ and its universal fluctuations $\mbox{var }g$:
\begin{eqnarray}\label{eq:varA}
\mbox{var }\A &=&\frac{\phi^2 \mbox{var }g}{(2+\phi^2)\mbox{var
}g+2(1+\phi^2)\left(\frac{C}{C_\mu}-1\right)^2g^2},\\
\mbox{var }\A&=&\frac{1}{1+2(g^2/\mbox{var }g)(C/C_\mu-1)^2 },\,\,
\phi\gg 1.\label{eq:highField}
\end{eqnarray}
The high-field limit $\phi\gg 1$ of var $\A$ is universal and Eq.
(\ref{eq:highField}) is valid for  {\em arbitrary} temperatures $t$,
as long as $T\ll\Thou$. However, this value is temperature-dependent
due to temperature dependence of UCF. As Eqs.
(\ref{eq:varA},\ref{eq:highField}) demonstrate, measurements of $\A$
could serve as a tool to find interaction strength $C_\mu/C$.
Recently an experiment in micrometer-sized Aharonov-Bohm (AB) rings
\cite{Bouchiat} found $\A\approx 0.3$ analyzing the amplitude of the
AB-oscillations in the non-linear conductance. This asymmetry was
found to be rather insensitive to variation of sample's conductance
$g\gg 1$, which means that the 2nd term in the denominator of Eq.
(\ref{eq:highField}) is not large and interactions are rather
strong, $C_\mu/C\approx 0.9$.

Since the asymmetry parameter $\A$ is not sensitive to the
fluctuations of the density of states, it probes interactions more
directly than $\Ga$. The symmetrized $\Gs$ is measured in
\cite{Zumbuhl} on the background of UCF in the linear conductance,
so that the interaction strength could be evaluated only from the
amplitude of $\Ga$. Similar to \cite{Bouchiat} measurement of $\A$
in quantum dots would provide a more precise measure of interactions
then those of $\Ga,\Gs$, which we consider below.

For pair correlators of $\Ga$ and $\Gs$ all fluctuating quantities
of Eq. (\ref{eq:defG}) should be taken into account:
\begin{eqnarray}
\label{eq:main}
\binom{\langle\Ga(t,\phi)\Ga(t',\phi')\rangle}{\langle\Gs(t,\phi)\Gs(t',\phi')\rangle}
&=&\binom{{\mathcal F}_-}{{\mathcal F}_++\frac{N^2}{2N_1N_2}
\left(\frac{C}{C_\mu}-\frac{2N_1}{N}\right)^2} \nonumber\\
&\times& \left(\frac{2\pi}
{\Delta}\frac{C_\m}{C}\right)^2\frac{N_1^3 N_2^3}{N^6}{\mathcal F}_+
.
\end{eqnarray}
Some properties of these correlators at $t=t',\phi=\phi'$ were
discussed in Ref. \onlinecite{PB}. Equation (\ref{eq:main}) allows
one to easily evaluate fluctuations in the non-linear conductance as
a function of temperature and flux. For example, at sufficiently
high flux difference, correlation functions decay strongly.
Similarly to 'magnetic fingerprints' in the linear conductance
\cite{meso}, the nonlinear conductance is randomized beyond
$|\phi-\phi'|\sim 1$.

We note some similarities between Eqs. (\ref{eq:AA}) and
Eq.(\ref{eq:main}). Despite these similarities, an important
difference is that correlations of ${\A}=\Ga/\Gs$ (see
(\ref{eq:AA})) are not given by a simple ratio of its averaged
numerator and denominator of Eq. (\ref{eq:main}),
$\la{\A}^2\ra\neq\la\Ga^2\ra/\la\Gs^2\ra$. In contrast, to leading
order in $1/N\ll 1$ $\Ga$ and $\Gs$ are uncorrelated.

{\em Weakly interacting limit.} In the limit of weak interactions
$C_\mu/C\to 0$ and the magnetic field asymmetry in Eq.
(\ref{eq:main}) vanishes. However, even without interactions
non-linear conductance exists \cite{AK,KL}. The energy scale
$eV_c\sim\hbar/\dwell$ was argued to be the only energy scale for
the crossover from linear to non-linear transport. In the open
diffusive samples considered in \cite{AK,KL} $eV_c\sim\Thou$ since
$\dwell\sim \erg$.

Equation (\ref{eq:main}) helps us to relate the correlations of
non-linear conductances at arbitrary $t,\phi$ and $t',\phi'$ in
weakly interacting dots to those of the {\em linear} conductance. In
particular, we find that in this limit the ratio of correlation
functions of linear and non-linear conductances is constant at fixed
$\dwell$. Particularly, for UCF and fluctuations of $\G$, we find a
universal relation
\begin{eqnarray}\label{eq:weaklimit}
\mbox{var } g=2(\hbar/\dwell)^2\mbox{ var }\G_s
.\label{eq:GG}
\end{eqnarray}
 insensitive to temperature $t$ and magnetic
flux $\phi$. This relation illustrates the qualitative arguments of
\cite{AK,KL} that transport coefficients in the non-interacting
system scale with $\hbar/\dwell=N \Delta/2\pi$.

In the crossover from weak to strong interactions a typical
amplitude of the fluctuations in $\Gs$ changes from $\d\Gs\sim
1/(N\Delta), \d\Ga\approx 0$, see Eq. (\ref{eq:weaklimit}), to a
value $\d\Gs, \d\Ga\sim 1/(N^2\Delta)\cdot\mbox{min }
\{1,\sqrt{1/t}\}$. Therefore, in the experiment, the interaction
induced scaling of a non-linear response $\d\Ga\sim \d\Gs$
proportional to $1/N^2$ can be observed. In contrast for a
non-interacting system we predict (see Eq. (\ref{eq:weaklimit}))
that $\Ga=0$ and $\Gs$ is proportional to $1/N$. Indeed, in the
experiment by Angers \etal  \cite{Bouchiat}, this observation allows
one to evaluate the interaction strength in Aharonov-Bohm rings,
$C_\mu/C\approx 0.9$.

\subsection{Multi-terminal dot}\label{multit}
We point out that similarly to two-terminal dots one can consider
multi-terminal dots as well. Since $\Ga$ is very sensitive to
interactions and is often easier to find experimentally, we
concentrate on a similar anti-symmetric component $\G_{a,\a\a\a}$
and on a {\em mixed} non-linear conductance $\G_{a,\a\b 0}$, a
derivative with respect to the gate voltage $V_0$ and contact
voltage $V_\b$.

One can demonstrate that in two-terminal dots the non-linearity with
respect to $V_0$ is absent. Indeed, $\G_{\a\b 0}$ in Eq.
(\ref{eq:Gabg}) is only due to $u_0(\phi) g'_{\a\b}(\phi)$. The
derivative $u_0$ is always field-symmetric, see Eq. (\ref{eq:u0}).
In two-terminal dots $g'_{\a\b}(\phi)$ is symmetric too,
$g'_{\a\b}(\phi)=g'_{\a\b}(-\phi)$. As a consequence $\G_{\a\b 0}$
is even in magnetic field {\em for two-terminal dot} (the classical
effect of $V_0$ in two-terminal samples was considered
 in \cite{Rikken_Wyder}). However, such asymmetry appears in a
multi-terminal set-up, since $g'_{\a\b}(\phi)\neq g'_{\a\b}(-\phi)$.

We find for the variances of $\G_{a,\a\a\a}$ and $\G_{a,\a\b 0}$,
\begin{eqnarray}
\mbox{var }\G_{a,\a\a\a}&=&\left(\frac{2\pi}
{\Delta}\frac{C_\m}{C}\right)^2\frac{N_\a^3(N-N_\a)^3}{N^6}{\mathcal
F}_+{\mathcal F}_-,\\
 \mbox{var }\G_{a,\a\b 0}&=&\left(\frac{\pi C_\mu}{\nu_s e^2}
 \right)^2\frac{N_\a^2 N_\b^2}{N^4}{\mathcal F}_-.\label{eq:ab0}
\end{eqnarray}
Equation (\ref{eq:ab0}) is of particular importance since it can
serve to find $C_\m$ in an experiment. This result does not contain
an apriori unknown $C$. It is useful to notice that for strongly
interacting electrons the effect of gate voltages is weak ($C\to 0$)
and therefore voltage difference between gates and the dot can be
rather large. Therefore, the non-linear conductance becomes
insensitive to the gate voltage and scales with $C_\mu\approx C\to
0$. However, for weakly interacting electrons $C_\mu\to \nu_s
e^2/\Delta$, the internal potential closely follows the gate voltage
and so the effect of gates becomes measurable.

\section{Discussion}\label{discussion}

The results we have obtained can now be compared with results for
the linear conductance. From such a comparison we can assert that it
is indeed the weakly non-linear transport regime which should be
used to extract information on the interaction strength rather than
the linear conductance. At finite temperature $t=2\pi T/N\Delta$, in
open quantum dots with many ballistic channels, $N\gg 1$, the
amplitude of the random linear conductance fluctuations calculated
in the strong interaction limit $e^2/C\Delta\gg 1$ depends on
interactions as $\d g\sim -t^2$, for $t\ll 1$ and $\d g\sim
-1/Nt^{3/2}$ for $t\gg 1$ \cite{BLF}. On the other hand, random
fluctuations of the non-linear conductances $\Ga,\Gs$ in this limit
$e^2/C\Delta\gg 1$ are $\propto 1/(N^2\Delta)\mbox{ min }\{1,
1/t\}$. For a voltage of the order of $eV\propto N\Delta$, the
fluctuating contribution of the non-linear component to the current
is $\propto \G (eV/\Delta)^2 $ and can be compared with the
fluctuating linear contribution $\d g\, eV/\Delta$. We conclude that
for voltages in the interval $\mbox{ min }\{N t^2, 1/\sqrt{t}\}\ll
eV/N\Delta\ll 1$ the transport is still weakly non-linear, but the
quantum contribution due to  Coulomb interaction to non-linear
transport is stronger then the linear contribution. This estimate
demonstrates that in dc-transport the effect of Coulomb interactions
in open mesoscopic conductors is best investigated in the non-linear
regime.

Another important issue is the sensitivity to small magnetic flux,
$\phi\ll 1$, when $I(\phi)$ in a two-terminal dot becomes an asymmetric
function due to Coulomb interaction. In the experiment $t\gtrsim
0.5$ \cite{Zumbuhl}, so temperature should be taken into account.
Although we do not consider inelastic scattering due to
$T\neq 0$, the non-linear conductance $\Ga(T)$ is temperature dependent since scattering is energy dependent. Of particular interest for the experiment \cite{Zumbuhl}
is the quantity $g_{B-}=(2\nu_s e^3/h)V\Ga$ in the limit where $\Ga(\phi)$ is still linear
in flux, $\phi\lesssim 1$, see Fig. \ref{fig1b}.  We use Eq. (\ref{eq:main}) and find
\begin{eqnarray}\label{eq:expfit}
\mbox{rms }\Ga(\phi) &=& \left(\frac{2\pi}
{\Delta}\frac{C_\m}{C}\right)\left(\frac{N_1^3
N_2^3}{N^{10}}\right)^{1/2}\cdot 4\phi\nonumber\\
&\times& \left( \int\frac{\pi^4 t^4x^2y^2(1+y)
e^{-x-y}dxdy}{2\sinh^2 \pi t y\sinh^2{\pi t x}} \right)^{1/2}
\end{eqnarray}
The data obtained at $T= 4\mu eV\ll\Delta=7\mu eV$ were fitted to a
combination of the zero-temperature results \cite{SB,SZ}
corresponding to a $\mbox{rms }\Ga$ given by the first line of
(\ref{eq:expfit}) with $4\phi$ substituted by $\Phi/\Phi_0$
\cite{Zumbuhl}.

It is noteworthy that a treatment of the crossover leads instead to
Eq. (\ref{eq:expfit}) with a non-linear conductance $\mbox{rms
}\Ga\propto \phi$, where $\phi\propto \Phi/\Phi_c$ is defined by Eq.
(\ref{eq:Flux}), and not $\Phi/\Phi_0$. Indeed, the flux dependence
of $\Ga(\Phi)$ starts to saturate at a scale which is parametrically
smaller then $\Phi_0$. This is a consequence of the existence of two
relevant time-scales, $\dwell$ and $\erg$. Thus an attempt to fit
the data to a theory of open diffusive samples \cite{SZ} by
substituting $\erg\to\dwell$ fails to catch the difference between
$\Phi_0$ and $\Phi_c\propto \Phi_0(\erg/\dwell)^{1/2}$. Being an
order-of-magnitude estimate to ballistic quantum dot results, the
substitution $\Phi_c\to\Phi_0$ misses parametric difference between
these fields due to the zero-dimensional physics of a chaotic dot.
This is especially important if the dependence of the effect on $N$
is considered, and $\mbox{rms }\Ga\propto N^{-5/2}$  at $T\to 0$.

We also point out that the temperature effect should have appeared
even at such low temperatures as $T=4\mu$eV. Indeed, the last factor
of Eq. (\ref{eq:expfit}) is equal to 1 for $t\ll 1$ and
$\sqrt{2}\pi/12t$ at $t\gg 1$. Because of exponential suppression of
the integrands at $x,y\sim 1/(\pi t)$, for $t\gtrsim 0.5$ the high
temperature asymptotic is a good approximation. Since $\Phi_c\propto
\sqrt{N}$, the nonlinear conductance as a function of $N$ and $T$
behaves as $\mbox{rms } \Ga\propto N^{-3/2}T^{-1}$.

To summarize, we discussed the symmetric and anti-symmetric
components of non-linear conductance through chaotic two-terminal
and multi-terminal quantum dots. Their correlations were found to
strongly depend on the interplay between the asymmetry of ballistic
contacts and the strength of Coulomb interactions. We discussed the
effect of a variation of the gate-voltage on the anti-symmetric
component of weakly non-linear conductance through multi-terminal
dots. We investigated the conditions on sample contacts, magnetic
field and temperature under which the (relative) magnetic field
asymmetry is maximal and thus most easy to detect.

We thank David S\'anchez for his valuable comments. We acknowledge
discussions with Piet Brouwer, Dominik Zumb\"uhl, and H\'el\`ene
Bouchiat. This work was supported by the Swiss National Science
Foundation, the Swiss Center for Excellence MaNEP and the STREP
project SUBTLE.

\end{document}